# It is Free and Always Will Be

TRADING PERSONAL INFORMATION AND PRIVACY
FOR THE CONVENIENCE OF ONLINE SERVICES

BY BRANDON ADAMS, AARON CLARK, AND JOSH CRAVEN

# Table of Contents

# Abstract

Internet users today are constantly giving away their personal information and privacy through social media, tracking cookies, "free" email, and single sign-on authentication in order to access convenient online services. Unfortunately, the elected officials who are supposed to be regulating these technologies often know less about informed consent and data ownership than the users themselves. This is why without changes, internet users may continue to be exploited by companies offering free and convenient online services.

# Introduction

"It's free and always will be" is the slogan visitors see on the signup page for Facebook. However, at what cost are "free" services like social media and lifetime email? Who really owns the data you upload to the cloud? Who is responsible when third parties access the private data of millions of people who have never even heard of them? Are the elected officials currently in office knowledgeable enough about current technology to write and pass legislation that will protect consumers of these services? This paper will address those questions and provide more details about the problem.

After it became known that "a political firm hired by the Trump campaign acquired access to private data on millions of Facebook users" (Granville, 2018), Mark Zuckerberg, the founder and current CEO of Facebook, was called to testify before the United States Senate on April 11 and 12, 2018.

More revealing than Zuckerberg's responses were the questions the Senators asked. At one point, Sen. Orrin Hatch, an 84-year-old Utah Republican set to retire this year, asked Zuckerberg how Facebook plans to stay free for its users. "Senator, we run ads," Zuckerberg said flatly (Jenkins, 2018).

Sen. Lindsey Graham, R-South Carolina, said he did not understand Facebook's terms of service, even though he is a lawyer (Wolverton, 2018). Finally, Ted Cruz, R-Texas, asked whether Facebook was "a First Amendment speaker expressing your views or are you a neutral public forum allowing everyone to speak?" Clearly, Cruz did not understand that Facebook is not responsible for what users post to its platform.

Disturbingly, the national leaders who are charged with protecting US consumers by regulating companies like Facebook revealed that they did not understand the technology. Users will thus continue to be vulnerable to exploitation by companies like Facebook, who use their data for profit, as well as a wide range of services and technologies today that users are forfeiting their privacy to use many different services besides just Facebook. In the meantime, people around the world struggle to understand concepts like data ownership and informed consent for tracking cookies, so-called free email, and single sign-on services – and typically use the services regardless.

## Amazon and Cookies

Facebook is not alone in its practices. Users entrust online businesses with personal information so they can purchase products online, connect with old friends and family members, and communicate with one another, among other things. One of the main methods for companies to collect personal information online is tracking cookies. According to Symantec, one of the largest U.S. cybersecurity software firms:

> *[A] tracking cookie is a detection for pieces of information stored on computers after visiting certain Web sites. Tracking cookies are distributed and retrieved across multiple Web site domains, allowing Web sites to monitor visitors' surfing habits. Some cookies can contain personal information or are bound to user profiles.* (Symantec, 2014)

Amazon is one the biggest online companies that uses tracking cookies to collect and monitor personal information. According to their privacy notice (Amazon, Amazon Privacy Notice, 2018), Amazon admits to transferring cookies on people's devices so customers can use specific features such one-click ordering and personalized advertisements. Amazon also states, "Cookies enable us to learn about what ads you see, what ads you click, and other actions you take on our sites and other sites (Amazon, Interest-Based Ads, 2018).

Amazon not only tracks users visiting the Amazon website: they also track users when they are visiting other websites. This information allows Amazon to tailor specific ads to their users and generate more income for themselves and third parties. Financially, tailoring targeted ads makes a great deal of sense for any online business. User-specific tailored ads lead to more purchases and greater revenue for all businesses involved. However this practice raises the question of when gathering this much personal information becomes too much. It is not clear that the information is safe in the hands of the businesses users put their trust in, or whether Amazon and other businesses can trust the third parties they collaborate with in the use of all this information. As we can see from the fallout of the recent Facebook/Cambridge Analytica scandal, trust is not always warranted.

Amazon has significant online market power. As of December 2017, Amazon averaged 197 million visits by United States users per month (Statista, 2017). That was 79 million more than second-place Walmart and equates to more than half the US population. Yet there is currently no federal or state government regulation regarding cookies or the amount of information a business can gather about a user's web activity. Amazon does allow users to opt out of personalized advertisements, but these advertisements are the default, rather than being a service that users can choose by opting in. This is a good example of how users must be vigilant about allowing businesses to access their personal information, including web-browsing activity.

## "Free" Email

Even if a user does not make online purchases on amazon, they most likely have a free email account somewhere. In addition, email has been around much longer than Facebook and Amazon. Yahoo mail began its service in 1997, with Google following suit in 2004. In 2016, email will connect more than 2.6 billion people worldwide and the numbers are projected to continue to rise (The Radicati Group, Inc., 2016). This high volume of traffic offers businesses a major opportunity to make use of communications data for their own profit. In 2017, Google curtailed its practice of scanning users' emails for targeted advertisement purposes, although they stated they would still scan for potential spam and phishing attacks. Google did so in an attempt to ease users' concerns about their data as the company entered the Internet infrastructure market (Wakabayashi, 2017). Yahoo and AOL, on the other hand, recently updated their terms of service to reiterate their intent to scan emails for business purposes (Hollister, 2018). Like Amazon, these businesses describe the purpose of this practice as enabling them to target specific ads of interest to the user. This is not always the case, however. For example, these companies have shared user emails with governments worldwide in order to crack down on illegal or undesirable activity. In the case of journalist Shi Tao of China, Yahoo provided records to Chinese state security officials that helped them convict Shi Tao of leaking state secrets (Kahn, 2005). The records that Yahoo provided showed Shi Tao's location, precisely when he emailed a particular memorandum, and the email's addressee, ultimately enabling security officials to convict the journalist. Whether or not one believes that Shi Tao is innocent, the story serves as a stark reminder that the user is using a "free" service where the information received and sent is never private; nevertheless, 2.6 billion people worldwide use these services to communicate, plan, and discuss personal and confidential information.

Regardless of the method used to collect this private data, the common theme is targeted advertising, a highly lucrative business practice that is not likely to go away: for example, Facebook alone generated $40 billion in advertising revenue last year (Wood, 2018). With this level of increasing annual revenue at stake, it is hardly surprising that businesses are working diligently to procure this information. It also makes sense that users choose to use these services to increase their quality of life, stay connected with family, use email for personal business purposes, or purchase products easily online. Businesses need the users, and the users of today's technological world have come to depend on these online businesses. Where we go from here depends on both the users and the businesses. The recent Facebook/Cambridge Analytical scandal has raised the issue of personal information being used for more than just targeted advertising, and may drive users to confront businesses sooner rather than later.

## Single Sign-On

Single sign-on authentication (SSO) is another web application that offers users convenience for the price of their personal information. Amazon, Google, and Facebook all take advantage of this application to provide greater ease of use to their users. The single sign-on authentication process allows a user to access multiple applications with one set of login credentials (Techopedia, 2018). For instance, rather than setting up multiple accounts for all the online periodicals they read, like *The New York Times*, *Washington Post*, and *Wall Street Journal*, a user can create each account using their own Google or Facebook account for authentication, reducing the number of user IDs and passwords they must remember and encompassing all the accounts into one. But like any "free" service, this comes at a price.

Most people do not give single sign-on a second look. Users must enjoy the ease of use that it offers, or SSO would not be as popular as it is today. What users do not generally know, however, is the amount of information that Google and Facebook are able to gather from the use of single sign-on. Using single sign-on allows these companies to gather a user's location (IP address), search history, apps used, events attended, photos and files (even deleted ones), and search history, and provides them access to the user's camera and microphone. According to data consultant and web developer Dylan Curran, the amount of data that Google and Facebook had specifically on him was large enough to fill 3.5 million Word documents (Curran, 2018). Similarly, the amount of information that these big businesses are gathering on their users is staggering. Some might argue that this information is harmless, and that  advertisement profiles of every customer (Curran, 2018) enable companies like Google to provide a better tailored online experience, rather than confronting users with advertisements that don't interest them. Others would argue that this information, while safe in the hands of advertisers, could become malicious when in the hands of companies that use it for other purposes than sales, like Cambridge Analytica.

So in the case of single sign-on and advertising profiles, again, users are forced to confront the idea of whether businesses are safeguarding their personal information, with little to no regulation. Can a business, in turn, trust the third-party businesses that it shares user information with? The amount of information that users allow businesses to have is staggering and may seem harmless on the surface; however, going forward it is imperative that both the business and the user be more mindful than they generally are today. Targeted advertising may be too lucrative for businesses to drop the model, but the importance of sharing this information securely with third parties cannot be understated.

## Cambridge Analytica and US Politics

Although the question of who really owns your personal information online is not new, it has gained attention worldwide recently since the alleged interference of analytics company Cambridge Analytica and their involvement in the 2016 presidential election. According to their mission statement, Cambridge Analytica's primary focus is "To deliver data-driven behavioral change by understanding what motivates the individual and engaging with target audiences in ways that move them to action" (Cambridge Analytica, 2018). The company works openly with public and privately held companies, political parties, and marketing firms nationwide. During the 2016 presidential campaign, candidate Donald Trump and his platform utilized Cambridge Analytica services. The research that Cambridge Analytica undertakes is not, in and of itself, illegal or improper. The moral and ethical questions raised by their actions center on the particular data they were analyzing on behalf of the Trump campaign, and the way that data was obtained.

The story began when Dr. Aleksandr Kogan, a psychology professor at Cambridge University, agreed to do what the university as a whole would not: work with Cambridge Analytica to create an app that hosted a personality/political quiz that pulled users' private information from Facebook, exploiting a loophole in Facebook's system. This loophole (already corrected by Facebook) allowed personal information that users gave on the social media platform to be shared with and used by the owner of the app. Users of the app knew their information was being used because the app presented itself as gathering data for "academic research." The exploitation of the loophole came about when Dr. Kogan's app extended its reach to pull the same personal information from the app user's friends, without their knowledge. According to *The New York Times*, "Only about 270,000 users – those who [had the app] – had consented to having their data harvested…" However, "[Dr. Kogan] ultimately provided over 50 million raw profiles to [Cambridge Analytica]" (Granville, 2018). The graphic below, created by *The Guardian*, summarizes the process.

It is important to note that the Trump campaign was not the first campaign to utilize data analytics and mining to influence the outcome of an election, and the practice has not been exclusive to one political party. In his article in the MIT *Technology Review*, Sasha Issenberg points out that "The Obama 2012 campaign used data analytics and the experimental method to assemble a winning coalition vote by vote" (Issenberg, 2016). They accomplished this primarily through the use of a team of data scientists, a mobile app, and a lot of research similar to that of the Trump campaign four years later. The data was used in the exact same targeted way in both campaigns. The primary difference between the two parties was whether users gave consent for their data to be used during the process. During the Obama campaign, users downloaded the "Obama" app, and knew they were giving information directly to Obama campaigners. In the Trump campaign, users unknowingly gave away their friends' information, under the impression that it was for an educational project, not a political campaign.

## Data Ownership

Facebook, known for providing a platform that is free for anyone to use, pays for its infrastructure by generating revenue from user data, by selling targeted advertisements. Thousands of companies pay for Facebook's advertising service, from Fortune 500s to "Mom and Pop shops" down the street, and for many it is their primary means of reaching a specific, targeted audience. Facebook is not the only company that offers targeted messaging and research, either. Other social platforms, like Instagram and Twitter, follow suit, while online stores such as Walmart, Amazon, and Google track similar information. The ethical questions that emerge from this are complex: in particular, who owns the information that a user gives online – the user, or the company that oversees the platform? More importantly, what information is actually private? Currently, these questions and many others like them are under debate in business, IT, and governments alike.

The Cambridge Analytica case, due to the strong and personalized marketing campaigns derived from personal information, is one of the leading examples that is generating discussion around these questions. Leaving aside the fraudulent manner in which the data retrieval was presented, the data was nevertheless vulnerable to collection. This data was used to create "fake news" – that is, information not grounded in provable fact – which was then distributed to individuals based on the interests revealed by their data. By most individual standards, this is not morally or ethically right. Other examples of inappropriate data targeting include reported cases of racist ad campaigns, discrimination against personal beliefs, and improper targeting of minors (users under the age of 18). Although these types of transgressions tend to be outliers, they are all serious issues that derive from data handled by trusted platforms.

Do Facebook users really have ownership of their information? Mark Zuckerberg claims that because a user has the ability to upload the media or remove it at any time, *the user* owns it outright.

During his congressional questioning (PBS NewsHour, 2018), Zuckerberg described Facebook's philosophy around data ownership: if someone decides to upload a photo or log personal information on Facebook, that person has the freedom and ability to remove the material at any time they wish. In addition, that person can control which "friends" (a subjective term used for friends, family, and acquaintances alike) can see the information the user publishes. What Zuckerberg failed to address, however, was the fact that users have no control over how that information is used by Facebook and third parties in marketing and advertising campaigns. Furthermore, Facebook philosophy is not unique in this philosophy, and its practices are common across the industry.

## Informed Consent

Even though users have agreed to share their data, because of the Cambridge Analytica incident, many social media platforms have had to answer uncomfortable questions in recent days about who owns the data on their systems. Throughout Mark Zuckerberg's testimony about Facebook's involvement with the Cambridge Analytica data breach, one key element repeatedly arose, calling into question a key IT and business concept: informed consent.

Fundamentally, the ability of online platforms to reach individuals and provide a custom web experience – from ads to suggested articles and more – is based on this concept. According to Business Insider, informed consent is "the idea that such companies can do whatever they want to with consumer data, or structure their interactions with consumers just about any way they want, as long as they disclose those practices and conditions first" (Wolverton, 2018). While this sounds reasonable, the problem is that most consumers are unaware of how *exactly* their information is being used. With every online product – social profile, software, web service, etc. – there is almost certainly an end-user license agreement that goes with it. This is the legal agreement that states what users are volunteering their information for, in exchange for the service they are using. The challenge is that this information is not typically provided in terms that are easily understandable, as in-depth legal language is often used and seldom explained, and it is seldom fully read by the average user. Although everyone checks the box stating that they "agree with the terms of service," what they are actually agreeing to remains a mystery for most.

With usage terms remaining a mystery to most consumers, the burden of responsibility has fallen on the businesses to determine what best practices should be, and what the terms of informed consent are. Up until now, the US government has stayed out of businesses' way, allowing them to self-govern and – at least theoretically – hold themselves responsible for their actions. When establishing a business model, however, this can get tricky when not every customer understands what the agreement terms are. But if the customer is satisfied by the services businesses are providing at a low or zero cost to the consumer, then they should have the right to sacrifice their privacy and give up their personal information.

## Conclusion

Consumers use ostensibly free services that claim to make their lives easier and more convenient, but the price they ultimately pay is anything but free. Moreover, while most users do not fully understand concepts like data ownership or informed consent, US elected officials are often even less knowledgeable. Until the US government gets a grasp on the basics of the business model, value proposition, or financial incentives of online services, corporations may continue to exploit the personal information and data of US consumers and people all over the world.

Wood, M. The data economy: The role of advertising. Marketplace.org. April 11, 2018. Retrieved from https://www.marketplace.org/2018/04/11/world/data-economy-role-advertising